\documentclass[onecolumn,amsmath,amssymb]{qm2008_abs}
\usepackage{graphicx}
\usepackage{dcolumn}
\usepackage{bm}
\begin{document}

\title{\Large The p/${\rm \pi}$  ratio $p_{T}$-dependence \\ in the RHIC range of baryo-chemical potential}
\bigskip
\bigskip

\author{\large N. Katry\'nska$^{\rm a}$, P. Staszel$^{\rm a}$ \\ for the BRAHMS Collaboration}
\email{n.katrynska@if.uj.edu.pl, ufstaszel@if.uj.edu.pl}
\affiliation{$^{\rm a}$Smoluchowski Institute of Physics, Jagiellonian
  University, \\ Krakow, 30-059, Poland}
\bigskip
\bigskip


\begin{abstract}
\leftskip1.0cm
\rightskip1.0cm
The BRAHMS measurement of proton-to-pion ratios in Au+Au and p+p
collisions at
$\sqrt{s_{NN}}$ = 62.4 GeV and $\sqrt{s_{NN}}$ = 200 GeV is 
presented as a function of transverse momentum and collision centrality
within the pseudorapidity range 0 $\leq
\eta \leq$ 3. The baryo-chemical
potential, $\mu_{B}$, for the indicated data spans from $\mu_{B}$
$\approx$ 26 MeV ($\sqrt{s_{NN}}$ = 200 GeV, $\eta=0$) to  $\mu_{B}$
$\approx$ 260 MeV ($\sqrt{s_{NN}}$ = 62.4 GeV, $\eta \approx 3$)
\cite{ionutqm}. The  p/${\rm \pi}$ ratio measured for Au+Au system at $\sqrt{s_{NN}}$ =
62.4 GeV, $\eta \approx 3$ reaches astounding value of 8-10 at 
$p_{T} \geq$ 1.5 GeV/c. For these energy and pseudorapidity interval no
centrality dependency of p/${\rm \pi}$ ratio is observed. Moreover, the
baryon-to-meson ratio of nucleus-nucleus data are consistent with
results obtained for p+p interactions.    
\end{abstract}

\maketitle

\section{Introduction}
In the last
decades, the intense theoretical and experimental investigations of the QCD phase
diagram in the regime of partonic and hadronic gas phases  led us to 
the picture depicted in Fig. 1. The dashed-dotted
red (online) line represents the crossover from Quark Gluon Plasma to
the hadronic state procured from the lattice QCD calculation
\cite{stephanov}. 
The experiemntal measurement of hadronic species abundances allows us to 
outline the dotted blue (online)
line as the chemical freeze-out of the hadronic gas. It is
remarkable that at low baryo-chemical potential that two curves
overlap, albeit at large $\mu_{B}$ a significant gap
between the temperature of the transition from the partonic to the
hadronic phase, $T_{c}$, and the temperature of chemical freeze-out is
predicted. 

The data of elliptic flow \cite{elliptic_flow} and p/${\rm
  \pi}(p_{T})$ \cite{ejkm, qm2008}
 around midrapidity (low $\mu_{B}$) have shown that the final hadronic state   
remembers the partonic fluid features. This is reflected in constituent
quark scaling of $v_{2}$ and an enhancement of baryon-to-meson ratios
that scales with the size of the created systems (see \cite{ejkm}, Fig. 2). 

These results support the view of hadronization process driven by the
parton recombination \cite{hwa_67} with the negligible final state interactions
between produced hadrons.
From Fig. 1 one can conclude that at large $\mu_{B}$ this picture 
might be spoiled by the final state hadron interactions leading to 
the transition from parton recombination scheme to
the hydrodynamical description with the common velocity field of
baryons and mesons \cite{hirano, broniowski}.

\section{Experimental Layout and Analysis}
The BRAHMS detector setup \cite{brahms_det} consists of two movable, narrow
spectrometer arms: the Midrapidity Spectrometer which operates in the
polar angle interval from $90^{\circ} \leq \Theta \leq 30^{\circ}$ (that corresponds in
the pseudorapidity interval $0 \leq \eta \leq 1.3$) and the Forward Spectrometer that operates in the
polar angle range from $2.3^{\circ} \leq \Theta \leq 15^{\circ}$ ($2
\leq \eta \leq 4$). Additionally, BRAHMS setup consists of the global
detectors used to determine the overall particle multiplicity,
collision vertex and centrality.

The Midrapidity Spectrometer is composed of the single dipole magnet
(D5) placed between two TPCs which are made use of tracking. Particle
identification is based on the Cherenkov
detector (C4 - signs like in \cite{brahms_det}) and Time of Flight
Wall (TOFW) measurement.  

The front forward arm is composed of two Time Projection Chambers 
(constituted track recognition in a high multiplicity enviroment), 
the back part - of three Drift Chambers 
and in the aggregate deliver particle track segments with
high momentum resolution using three dipole magnets. To identify
low momentum particles the Time of Flight (hodoscope H2) is used. For higher
mometa, particle identification is provided via then Ring
Imaging Cherenkov detector, situated behind H2 hodoscope.         
  
For data analysis we assume that acceptance and tracking efficiency are canceled in the
baryon-to-meson ratio but the proton and pion yield have been
corrected for PID efficiency, interactions of emitted particles with the beampipe,
the spectrometers natural budget and for the decays in fly. 

\begin{figure}
\includegraphics[width=.78\textwidth,totalheight=0.45\textheight,angle=0]{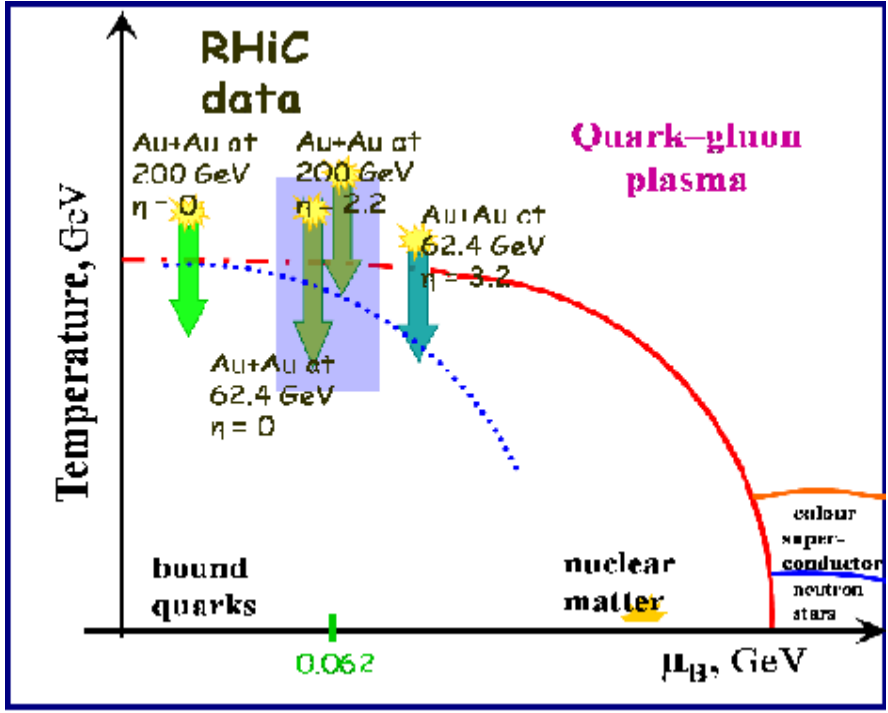}
\caption{The scheme of QCD phased diagram: the dashed-dotted red (online)
  line emblematizes the crossover between Quark Gluon Plasma and
  hadronic phase. The dotted blue (online) curve represents the
  chemical freeze-out.The arrows denote the value of baryo-chemical
  potential for definite colliding systems \cite{braun_m}.}
\end{figure}

\section{Results and discussion}

\begin{figure}
\includegraphics[width=0.9\textwidth,totalheight=0.33\textheight]{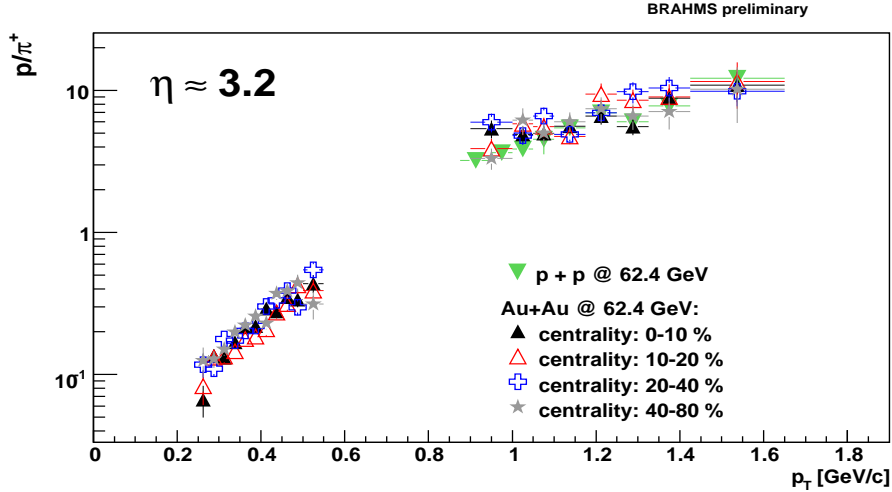}
\caption{Centrality dependent ratio of p/${\rm \pi}$ for Au+Au system
  at  $\sqrt{s_{NN}}$ = 62.4 GeV for $\eta$ = 3.2 in comparison with
  p+p collisions. The errors are only statistical.}
\end{figure}

It has been already shown that the p/${\rm \pi}$ ratios at the
intermediate $p_{T}$ range can vary very strongly depending on both
charge and pseudorapidity of indicated species, 
as well as on energy and size of colliding system. 
Fig. 2, \cite{ejkm}, presents the
p/${\rm \pi}(p_{T})$ ratio at midrapidity for Au+Au at $\sqrt{s_{NN}}$
= 200 GeV ($\mu_{B}$ = 26 MeV) in comparison with theoretical predictions based on parton
recombination model \cite{hwa_70} and hydrodynamical descritpion.
The characteristic growth at intermediate $p_{T}$
region, predicted by above-mentioned descriptions, seems to be more
consistent with the depiction of recombination
model. The hydrodynamic scenario, proposed in \cite{hirano}, describes
properly only the low momentum data associated mainly with the soft
component.

Fig. 2 compares the p/${\rm \pi}$ ratio from p+p and Au+Au collisions
at $\sqrt{s_{NN}}$ = 62.4 GeV and $\eta$ = 3.2 ($\mu_{B} \approx $ 250
MeV, \cite{ionutqm}). 
Unexpected high value of 10 at $p_{T}$ = 1.5 GeV/c of proton-to-meson ratio
\cite{hwa_76} is observed.  

There is remarkably little difference in the pi/p ratios from a very wide range of systems. 
This is in contrast to the trends at midrapidity and forward rapidity regimes for Au+Au at
$\sqrt{s_{NN}}$ = 200 GeV where significant medium effect of baryon-to-meson
ratios depending on system size is seen.   
However, it must be admitted that at
forward pseudorapidity a lot of protons come from beam fragmentation, 
that high value of proton-to-pion ratio  for all
intervals of centrality for nucleus-nucleus collisions is a puzzle
indicating that mechanism of baryons-to-mesons production is taking
place rather on elementary interaction domain.

Fig. 3 presents the Au+Au collisions for $\eta$ =
0.0 at $\sqrt{s_{NN}}$ = 62.4 GeV marked with open red (online)
triangles and the Au+Au reactions for  $\eta$ = 2.2 at
$\sqrt{s_{NN}}$ = 200 GeV marked with the black triangles. The
selection of pseudorapidity intervals namely $\eta$ =
0.0 for Au+Au @ 62.4 GeV and $\eta$ = 2.2 for Au+Au @ 200 GeV allow   
to obtain overlap in ${\rm \bar{p}/p}$, thus  $\mu_{B}$, for the 
observed phase space of
system at various energies. As indicated in Fig. 1
the shown data are measured at $\mu^{Au+Au @ 200
  GeV}_{B}$= $\mu^{Au+Au @ 62.4 GeV}_{B}$ = 62 MeV.
Considerably lower value depicted by grey
stars displays the p/${\rm
  \pi}$ ratio for p+p system at $\sqrt{s_{NN}}$ = 200 GeV. The
astonishing conformity proton-to-pion ratios for collate
heavy ions collisions evidences
that the baryon and meson production at the covered $p_{T}$ interval
is dominated by medium effects. These effects can be seen throughout
the observed enhancement of p/${\rm \pi}(p_{T})$ for nucleus-nucleus
systems with reference to the results for elementary interactions.
The data infer possible scaling of
baryon-to-meson ratio with baryo-chemical potential for dense systems.
In addition it is presented the comparison with THERMINATOR model      
with well-defined collective expansion and successive evaporation of hadrons from the
hypersurface of the fireball \cite{broniowski}.  

\begin{figure}
\includegraphics[width=0.9\textwidth,totalheight=0.33\textheight]{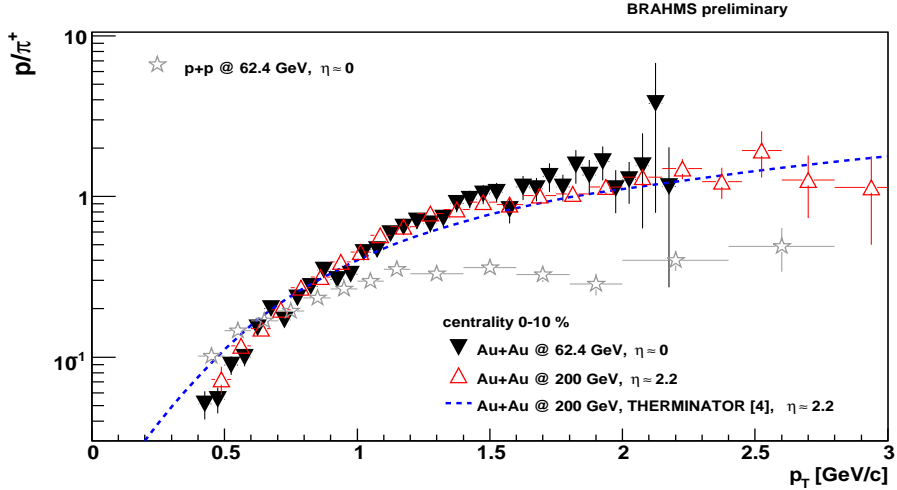}
\caption{Proton-to-pion ratio for Au+Au (0-10\% centrality) and p+p
  collisions measured at the same
  value of baryo-chemical potential $\mu_{B}$ = 62 MeV. The errors are
only statistical.}
\end{figure}

\section{Summary}

Concluding, the BRAHMS experiment has presented the proton-to-pion ratios
in Au+Au and p+p collisions at $\sqrt{s_{NN}}$ = 62.4 GeV and
$\sqrt{s_{NN}}$ = 200 GeV as a function of transverse momentum and collision
centrality. As it has been shown the indicated data give the possibility of
studying baryon and meson production in the
wide range of baryo-chemical potential $\mu_{B}$.
For Au+Au and p+p measurement at
$\sqrt{s_{NN}}$ = 62.4 GeV for $\eta$ = 3.2 the astounding value of p/${\rm \pi}(p_{T})$
ratio has been noted. At these energy and pseudorapidity interval no
centrality dependency of p/${\rm \pi}$ ratio is observed. Furthermore, the
baryon-to-meson ratio of nucleus-nucleus data are consistent with
results obtained for elementary p+p reactions.   

To test the effect of baryo chemical potential we have compared $\eta$ = 2.2 and $\sqrt{s_{NN}}$ = 200 GeV with $\eta$ = 0.0 and $\sqrt{s_{NN}}$ = 62.4 GeV. The pi/p ratios are remarkably similar.

\end{document}